\documentclass[final,5p,times,twocolumn,number,sort&compress]{elsarticle}

\usepackage{blindtext}
\usepackage{slashed,multirow,relsize,soul}
\usepackage{feynmf}
\usepackage[normalem]{ulem}
\usepackage{color}
\usepackage{siunitx}
\usepackage{array}   
\usepackage{geometry}
\usepackage{longtable}
\newcolumntype{L}{>{$}l<{$}} 
\usepackage{bbold}
 \usepackage{mathrsfs}
\usepackage{braket}
\usepackage{physics}
\usepackage{graphicx}
\usepackage{xcolor}
\usepackage{amssymb}
\usepackage{lipsum}
\usepackage[capitalise]{cleveref}

\def\mphi{m_{\Phi}}
\def\mhiggs{m_{H}}
\def\lphi{\lambda_{\Phi}}
\def\lhiggs{\lambda_{H}}
\renewcommand{\phi}{\varphi}
\definecolor{darkblue}{rgb}{0,0,0.5}

\journal{Nuclear Physics B}

\begin{document}

\begin{frontmatter}



\title{From oversimplified to overlooked: the case for exploring Rich Dark Sectors}


\author[ift]{Asli Abdullahi}
\author[charles]{Francesco Costa}
\author[unibo,infn]{Andrea Giovanni De Marchi\fnref{lbnl}}
\author[unibo]{Alessandro Granelli}
\author[ncbj]{Jaime Hoefken-Zink}
\author[harvard]{Matheus Hostert}
\author[unibo,infn]{Michele Lucente\fnref{fnal}}
\author[unibo,infn]{Elina Merkel}
\author[unibo,infn]{Jacopo Nava}
\author[unibo,infn]{Silvia Pascoli}
\author[infn]{Salvador Rosauro-Alcaraz}
\author[unibo,infn]{Filippo Sala}

\affiliation[ift]{organization={Instituto de Física Teórica (IFT), UAM-CSIC},
            city={Madrid},
            country={Spain}}
            
\affiliation[charles]{organization={Institute of Particle and Nuclear Physics, Charles University},
            city={Prague},
            country={Czechia}}

\affiliation[unibo]{organization={Dipartimento di Fisica e Astronomia, Università di Bologna},
            addressline={via Irnerio 46},
            postcode={40126},
            city={Bologna},
            country={Italy}}

\affiliation[infn]{organization={INFN, Sezione di Bologna},
            addresslin={viale Berti Pichat 6/2},
            postcode={40127},
            city={Bologna},
            country={Italy}}

\affiliation[ncbj]{organization={National Centre for Nuclear Research (NCBJ)},
            addresslin={Pasteura 7},
            postcode={Warsaw, PL-02-093},
            city={Warsaw},
            country={Poland}}

\affiliation[harvard]{organization={Harvard University},
            city={Cambridge},
            country={USA}}

\fntext[lbnl]{Also visitor at: \emph{Berkeley Center for Theoretical Physics, Lawrence Berkeley National Laboratory, Berkeley, California 94720, USA}}
\fntext[fnal]{Also visitor at: \emph{Theoretical Physics Department, Fermi National Accelerator Laboratory, Batavia, Illinois 60510, USA}}

\begin{abstract}
The Standard Model (SM) of particle physics provides a very successful description of fundamental particles and their interactions but it is incomplete, as neutrino masses, dark matter and the baryon asymmetry of the Universe indicate.
In addition, the origin of masses and of the approximate fundamental symmetries call out for deeper explanations. The quest for a New SM Theory, that extends the SM to a more general theory, is ongoing. For decades the main focus has been on the TeV scale, but despite an impressive theoretical and experimental effort, no hints of new physics at such scale has been found in experiments. 

Dark sectors provide an interesting alternative to TeV scale extensions of the SM to explain the open questions in particle and astroparticle physics.
Going beyond minimal models, rich dark sectors extend the SM to a complex theory with multiple particles and interactions, in analogy to the SM itself. 
They have a wealth of theoretical and astrophysical/cosmological consequences and can lead to phenomenological signatures that can be markedly different to that of minimal ones.
These include short-lived particles and semi-visible decay signatures, as opposed to minimal models where new states are typically long-lived and purely visible or invisible resonances. Given the experimental configurations and analysis strategies, current dark sector
searches might miss such signatures. 
We advocate a dedicated programme of searches for rich dark sectors that overcomes the assumptions on minimality and on the long lifetime of particles and encompasses a broader range of possibilities. 
Here, we discuss a prototype model that includes a complex structure akin to the SM: multiple generations of fermions charged under a new spontaneously-broken gauge symmetry.

\end{abstract}

\begin{keyword}
dark sectors \sep experimental searches \sep dark matter \sep neutrinos \sep gravitational waves
\\\vspace{2ex}
Invited contribution to the Nuclear Physics B Special Issue on {\it Clarifying common misconceptions in high energy physics and cosmology}.
\end{keyword}

\end{frontmatter}

\section{Introduction}

Despite the great successes of the Standard Model of particle physics (SM), last but not least with the discovery of the Higgs boson in 2012 and the many precision tests, there is incontrovertible evidence that the SM is not the ultimate theory of particles and their interactions: non-zero neutrino masses, as required by neutrino oscillations, dark matter (DM), and the baryon asymmetry of the Universe cannot be explained within the SM.
In addition, the origin of masses and of the approximate symmetries of the SM call out for a deeper explanation.
The key aim of particle and astroparticle physics is to uncover the New Standard Theory that extends the SM to a (more) complete theory of fundamental particles and interactions. 

In this endeavour, a crucial question is: {\em What is the new energy scale?} 
We need guidance from theoretical considerations and experimental information. The range to be considered is very broad: neutrino masses can emerge from theories at energies as high as the grand unification scale, that provide an elegant framework but are inaccessible directly to us, or as low as the eV scale, 
implying a large hierarchy with that of the 
electroweak (EW) theory. 
Baryogenesis and leptogenesis have a similar spread in scales, with the requirement of having baryons by the time of big bang nucleosynthesis at MeV temperatures. Moreover, if the baryon asymmetry arises from a leptonic one, as in leptogenesis models, this needs to happen before sphalerons become inactive at the EW scale.  Dark matter allows an even larger breadth of scales: it could be as light as $10^{-22}~\mathrm{eV}$ if bosonic or keV if fermionic, and even heavier than the Planck mass, if it is not made of elementary particles. 

The past decades have witnessed an impressive theoretical and experimental effort in the exploration of the TeV scale, culminating in the LHC experiments and plans for future colliders. 
As no evidence for physics beyond the SM at the TeV scale has emerged, interest in the broader exploration of scales has been renewed.

\section{Dark sectors: extending the SM below the EW scale}

Dark sector (DS) models formalize the possibility of light new physics: 
they are models with new particles and interactions at scales below the EW one.
Experimental constraints imply that the connection with the SM needs to be feeble, hence the name “dark” or “hidden” or “feebly interacting”. This idea has gone from a fringe activity to the center stage in theory and experiments in recent years.
For an overview of DS, see, e.g., \cite{Agrawal:2021dbo} and \cite{Antel:2023hkf} and references therein.

Notable examples include the area of DM, in which the community has started looking beyond the “lamp post” of Weakly Interacting Massive Particles (WIMPs). Novel ideas in extremely light~\cite{Hui:2021tkt} as well as sub-GeV DM~\cite{Antel:2023hkf}  
have been proposed 
with candidates that can have properties significantly different from ``traditional'' WIMPs~\cite{Carr:2021bzv,Adams:2022pbo,Cirelli:2024ssz}. Experimentally, a strong effort is undergoing to extend the physics reach of current experiments to lower masses and to design new experiments targeting this interesting mass range~\cite{Cirelli:2024ssz}.

On the neutrino front, low-scale see-saw mechanisms~\cite{Mohapatra:1986bd,Gonzalez-Garcia:1988okv,Pilaftsis:1991ug,Deppisch:2004fa,Shaposhnikov:2006nn,Gavela:2009cd,Ibarra:2010xw} that can be realised with masses as low as the MeV-GeV scale~\cite{Asaka:2005an,Abada:2014vea} have attracted renewed interest thanks to their ability to explain neutrino masses, the baryon asymmetry~\cite{Akhmedov:1998qx,Asaka:2005pn,Drewes:2012ma,Abada:2015rta,Hernandez:2015wna,Abada:2018oly,Domcke:2020ety,Drewes:2021nqr,Hernandez:2022ivz} and dark matter~\cite{Dodelson:1993je,Shi:1998km,Asaka:2006nq,Laine:2008pg}, while being testable at experiments targeting heavy neutral leptons~\cite{Abdullahi:2022jlv}, charged lepton flavour violation and electroweak precision observables~\cite{Blennow:2023mqx}.
In these models, due to an underlying symmetry, neutrino masses can be naturally small and the mixing between active neutrinos and heavy fermions can be large and testable~\cite{Wyler:1982dd,Shaposhnikov:2006nn,Kersten:2007vk,Abada:2007ux,Gavela:2009cd,Moffat:2017feq}.
Such models typically accommodate a much wider number of possibilities for additional new physics, whether related to the DM puzzle or not. 

To cover the realm of possibilities for light new physics beyond the SM, the community has adopted a set of benchmarks that target generic couplings between the dark sector and the SM, often referred to as ``portals".
In this context, seesaw models for neutrino mass rely on a neutrino portal, i.e., a renormalizable interaction between the heavy neutral leptons and light neutrinos.
Another example is that of kinetic mixing between a dark photon and the SM hypercharge or photon, referred to as the vector portal.
The latter can mediate interactions of light dark matter fermions with SM particles and realize DM production through freeze-out or freeze-in.
Finally, in all generality, new scalar degrees of freedom can mix with the SM Higgs via dimension-3 or 4 operators, constituting the scalar portal.

While the neutrino, vector and scalar portals are renormalizable, another motivated possibility is that the portal between the dark sector and the SM is given by non-renormalizable operators.
In that case, the new physics is manifest at two separate energy scales: that of the dark sector particles, which by definition is accessible to experiments, and the higher new-physics scale which induces the effective operators connecting dark and SM particles.
This represents a hybrid approach to pursuing new physics and will serve as another example of the rich DS models we discuss below.

\begin{figure}[t]
    \centering
    \includegraphics[width=\linewidth]{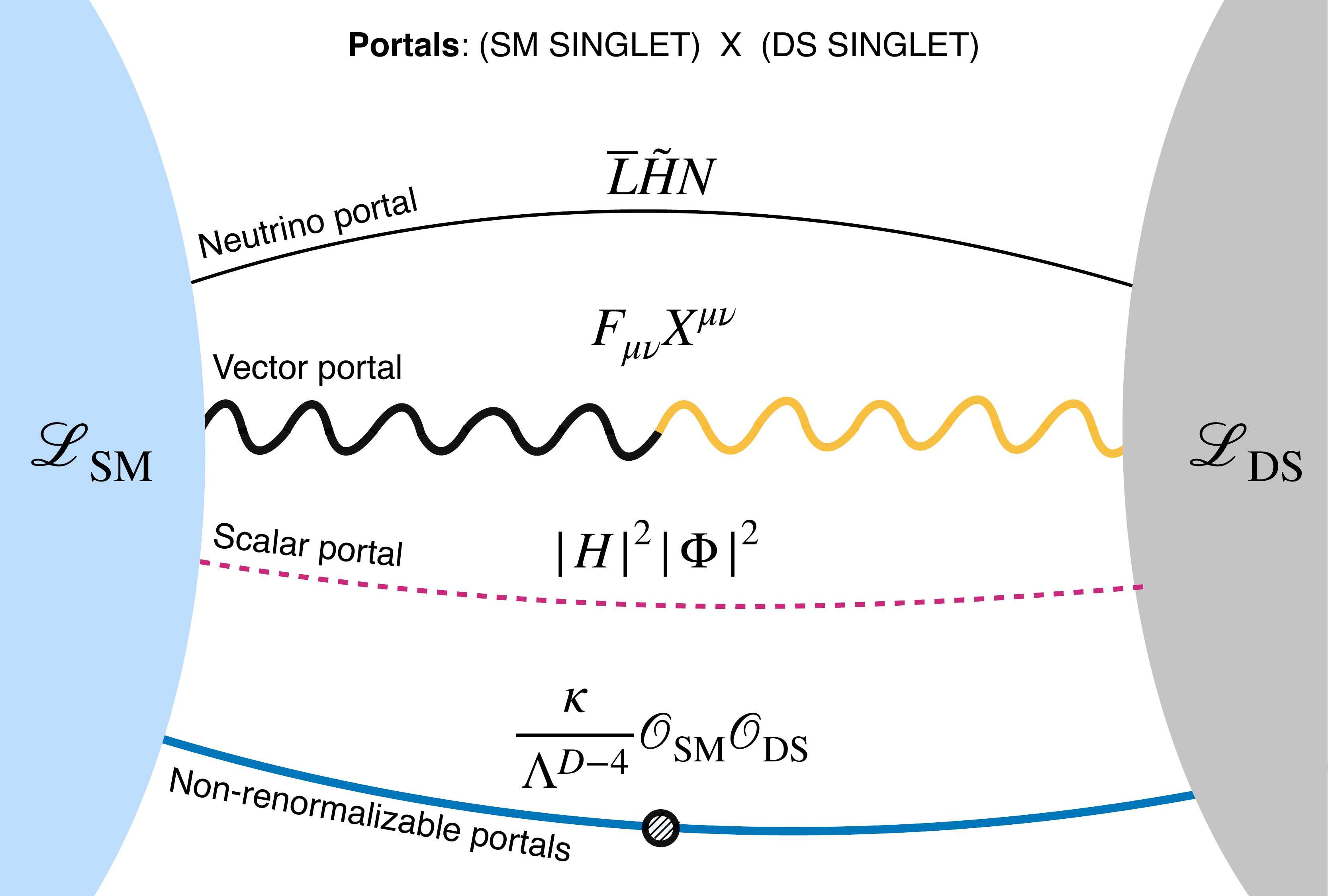}
    \caption{An illustration of the portal approach to searching for dark sectors at the high-intensity frontier.
    We include renormalizable portals as well as the possibility of non-renormalizable ones.
    \label{fig:3portal}
    }
\end{figure}

\subsection{Hints of dark sectors?} 
\label{sec:hints_DS}

One may ask whether current experimental data provides any hints for the existence of light particles. Indeed, some intriguing anomalies have been observed at low energies; however, the possibility that they are experimental artifacts cannot yet be ruled out.
One example includes the low-energy excess of electron-like events at MiniBooNE.
A reanalysis of MiniBooNE data increased the significance of this excess to over $4\sigma$~\cite{MiniBooNE:2020pnu}.
This anomaly is being tested at three liquid Argon experiments at the Booster Neutrino Beam (BNB) at Fermilab: MicroBooNE, ICARUS, and SBN~\cite{Machado:2019oxb}.
MicroBooNE has already performed several dedicated searches for exclusive SM processes like neutrino-induced $\Delta(1232) \to N \gamma$ radiative decays and neutrino coherent $\gamma$ production, ruling out several suspected backgrounds as the origin of the MiniBooNE excess~\cite{MicroBooNE:2021zai,MicroBooNE:2025qao,MicroBooNE:2025rsd}.
Curiously, when performing a more inclusive search, MicroBooNE reports instead a mild excess of photon-like events somewhat consistent with the MiniBooNE excess~\cite{MicroBooNE:2025ntu}. 
Although no specific process has been identified as the sole origin of this new excess, it provides an important step forward in resolving the MiniBooNE puzzle.

Other low-energy puzzles that have been studied within the context of DS models include the discrepancy between the $(g-2)_\mu$ measurement~\cite{Muong-2:2023cdq} and theoretical calculations based on the dispersive methods~\cite{Aoyama:2020ynm}, the discordance in $(g-2)_e$ and $\alpha_{\rm QED}$ measurements~\cite{Parker:2018vye,Morel:2020dww,Fan:2022eto}, the dimuon events at NuTeV~\cite{NuTeV:2000ehk}, a possible mono-photon excess at BaBar~\cite{BaBar:2017tiz} (see the discussion in \cite{Abdullahi:2023tyk}), the 17 MeV boson at ATOMKI~\cite{Krasznahorkay:2015iga,Krasznahorkay:2021joi,Krasznahorkay:2022pxs} and, more recently, at the PADME experiment~\cite{Nardi:2018cxi} (see the DS explanations in~\cite{Arias-Aragon:2024qji,Arias-Aragon:2025wdt,DiLuzio:2025ojt}).
While such anomalies may eventually find more mundane explanation elsewhere, many of them can easily fit within the context of DS.

Light particles could also lie at the origin of observational puzzles, in case their astrophysical or cosmological explanations become conclusively disfavoured. Examples include gravitational waves (GWs) at pulsar timing arrays (PTA), the 511 keV photons from the galactic bulge, and the Hubble tension~\cite{Riess:2019cxk} (for reviews see, e.g., \cite{Verde:2019ivm,DiValentino:2021izs})). 
In June 2023, PTA GW observatories have provided  evidence of a stochastic background of nanoHertz GW~\cite{NANOGrav:2023gor}. Its origin is unknown but falls into two main categories~\cite{NANOGrav:2023hvm}: an astrophysics one from mergers of supermassive black holes, or a cosmological one, with a supercooled sub-GeV-scale phase transition (PT) among the frontrunners, that requires multiple additional dark states~\cite{Bringmann:2023opz,Madge:2023dxc,Figueroa:2023zhu,Wu:2023hsa,Ellis:2023oxs,Ellis:2024xym}.

Since the 70's, INTEGRAL and other telescopes observe a photon line at 511 keV from the annihilation of $e^-e^+$ via para-positronium~\cite{Siegert:2015knp,Kierans:2019aqz}, from the galactic disk and bulge. While the disk component is understood from nucleosynthesis in stars, the origin of the bulge component is still under debate, see e.g.~\cite{Prantzos:2010wi,Fuller:2018ttb,Siegert:2021trw}. It could in fact be the first evidence for DM with mass of a few MeV, as first proposed in~\cite{Boehm:2003bt}, but also at and beyond ten MeV, as recently realised in~\cite{DelaTorreLuque:2024wfz,Aghaie:2025dgl}. All in all, the BSM explanation of this excess requires light (non-minimal) dark sectors, see e.g.~\cite{Ema:2020fit}.

\section{Rich dark sectors}

In the field of dark sectors, a typical guiding principle is that of minimality: extensions of the SM typically advocate one/few new particles and/or interactions, i.e. only one of the portals. Minimal models have the advantage of predictivity and allow for easier comparison between experimental bounds and sensitivities.
But the only known experimentally confirmed particle theory, a.k.a. the SM, is highly non minimal, with a complex gauge group, a scalar to break the EW symmetry, an additional mass scale generated by non-perturbative strong interactions, and a plethora of fermions, which even come in three generations. If analogy to the SM is used as a guiding principle, the DS should have new symmetries and multiple sectors, i.e. new (gauge and scalar) bosons and multiple dark fermions. We call them ``rich dark sector" models (RDSM).

In addition to the example of the SM, there are other key theoretical motivations for rich DS. The explanation of the multiple pieces of evidence of physics beyond SM (i.e. neutrino masses, leptonic mixing, DM and the baryon asymmetry) in a common picture requires multiple particles, at the very least multiple heavy neutral leptons. Typically, low scale see-saw models call for quasi-preserved lepton number, e.g. in inverse or linear see-saw, to avoid finetuning of the Yukawa couplings and large radiative corrections to light neutrino masses, and therefore assume multiple types of dark fermions. Even more so, the explanation of the new physics scale and in particular a dynamical origin for the mass scale of the dark particles requires a scalar that acquires a vacuum expectation value (vev)  via spontaneous symmetry breaking (SSB) or confined strong interactions within the dark sector components.
Coming to dark matter, models where it originates in a multicomponent dark sector are intensely studied, e.g. inelastic~\cite{Hall:1997ah,Tucker-Smith:2001myb} and secluded~\cite{Pospelov:2007mp} DM. Rich dark sector are also required by some production mechanisms motivating sub-GeV DM, like SIMPs~\cite{Hochberg:2014dra}, ELDERs~\cite{Kuflik:2015isi} and forbidden DM~\cite{DAgnolo:2015ujb}.
Finally on a phenomenological side, rich dark sectors are motivated by BSM explanations of existing experimental and observational anomalies, like those mentioned in Sec.~\ref{sec:hints_DS}.

Rich dark sectors have been identified as one of three “Big ideas” in RF6 Snowmass process in the US but the focus has been mainly on DM, e.g., inelastic DM, while their connection to HNLs and other dark particles has received little attention so far. 
Dark dynamics connected with the SM by non-renormalizeable portals similarly deserve further scrutiny, see~\cite{Contino:2020tix,Costa:2022pxv} for first attempts to characterize them systematically.

In many ways, RDSM have been around for a long time, often as the result of top-down theoretical considerations or as a consequence of a new confining sector of particles.
Some examples include Hidden Valleys~\cite{Strassler:2006im,Han:2007ae,Craig:2015pha}, mirror sectors~\cite{Kobzarev:1966qya,Foot:1991bp,Berezhiani:1995am}, or Twin Higgs models~\cite{Chacko:2005pe}.
In this article, we focus on a different perspective: namely, a bottom-up approach that aims to broaden the scope of phenomenological benchmarks from minimal to rich dark sector models.
Given the huge number of possibilities that can be explored, we narrow the discussion to RDSMs that most resemble existing DS benchmarks, multiplying the particle content in the dark sector and opening up more than one portal between SM and DS at a time.

\begin{figure}[t]
    \centering
    \includegraphics[width=\linewidth]{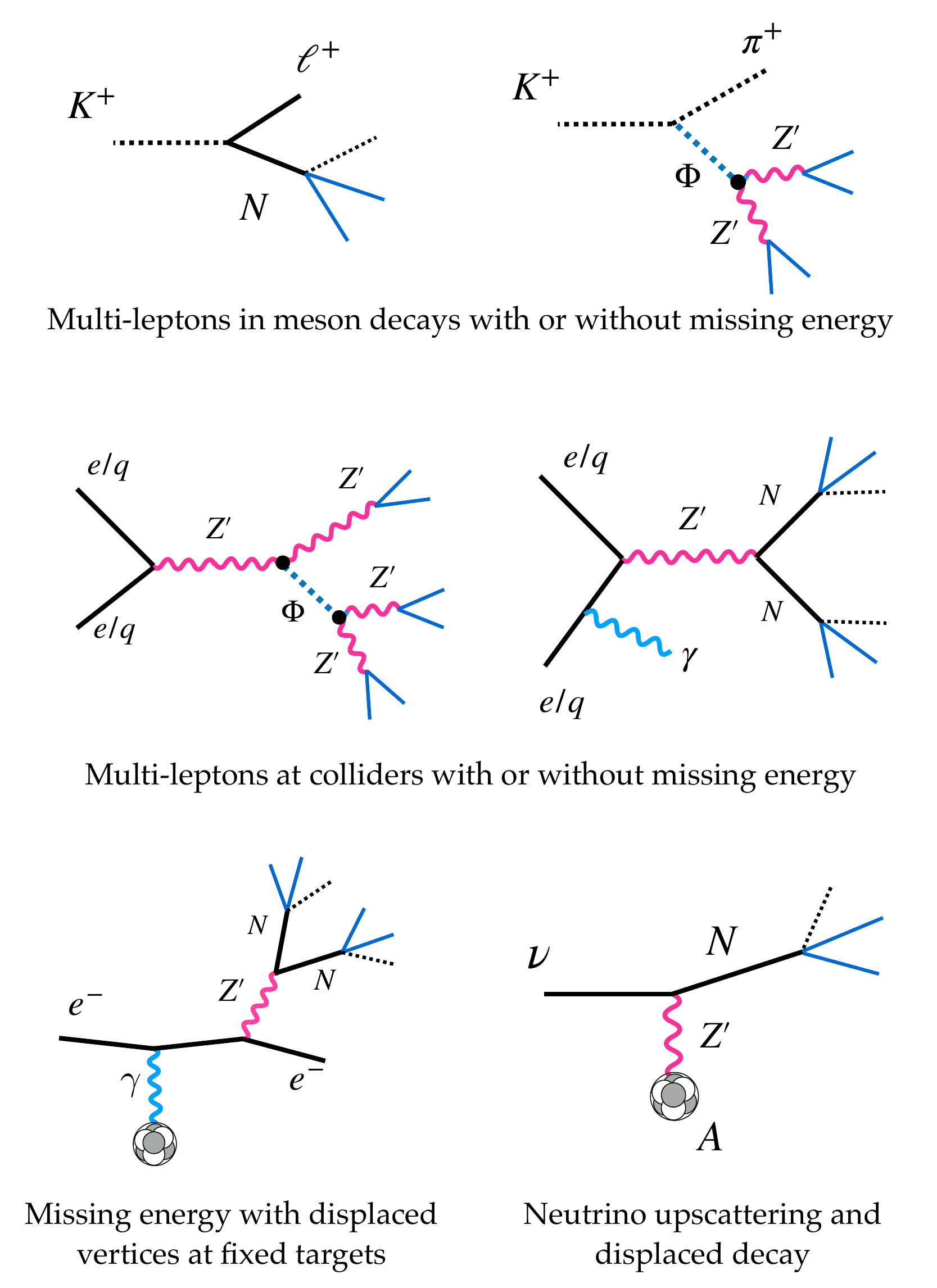}
    \caption{A few examples of experimental signatures, where dark blue lines represent some visible charged final states and dotted lines a neutrino or other invisible particles.
    For instance, in the top left diagram would represent decays such as $K^+ \to \mu^+ ( N\to \nu e^+e^-$).
    In rich dark sectors, the new particles often decay semi-visibly and are not targeted by simpler searches for visible or fully invisible resonances.
    \label{fig:3portal_examples}
    }
\end{figure}

\subsection{A unique phenomenology}
The main searches of DS go via the decays of dark photons, dark scalars and dark fermions into SM particles that can be detected in experiments. Thanks to their less simplistic features, like a multiportal connection to the SM and non-negligible couplings between multiple dark particles, rich dark sector models have signatures distinctively different from minimal models. 

For dark photons and dark scalars, there are mainly two types of searches that have been considered: i) visible searches in which the dark particle decays into a dilepton pair that exhibits a resonance in correspondence to its mass; and ii) invisible searches, in which the dark boson decays into invisible particles, e.g. DM or neutrinos, that appear as missing energy. This is the behaviour expected in minimal models, but it is not the case for rich ones. In the latter, the novel smoking gun signatures are fast {\em semi-visible decays} (i.e. that have both visible decay products and missing energy) with multiple leptons, and possibly decay chains.  
This has a strong impact on the experimental searches if they are not sufficiently inclusive. Cuts on invisible (or visible) energy would mistake the semivisible events as backgrounds. Visible and invisible dark photon bounds can be weakened by orders of magnitude as is discussed below. 

The other striking difference to minimal models, is that typically particles in rich dark sector models are shortlived  (SLP) (except, possibly, for the lightest one). This is due to the fact that there are multiple dark particles that can decay one into the others (plus SM ones) and whose couplings between them, without fine tuning, are expected to be quite large. This leads to lifetimes which can be shorter than in minimal models by many orders of magnitude. In minimal models, in fact, the dark particles can only couple to the SM via tiny couplings and, consequently, their lifetimes are expected to be very long, that is they are long lived (LLP). 

It should be pointed out that a large fraction of dark sector searches have focussed on long lived particles as expected in minimal DS models: this allows for dedicated searches and background reduction but may miss shortlived particles. For instance, this may be the case for experiments ``\`a-la beam dump” in which there is a significant distance between production, e.g. at a target, and subsequent decay in a far away detector. If the particles are short-lived, they may not reach the detector, having decayed earlier. Therefore, the bound obtained in minimal models would not apply as such and would need to be reconsidered, they may even be evaded completely.

We present below in Sec. \ref{sec:HNL@RDS} and \ref{sec:DP@RDS} a detailed discussion of the new phenomenology of dark particles and required search strategies. 

\subsection{A wealth of experimental opportunities.} 
The search for SLPs and semi-visible signatures in MeV--GeV dark sectors has received relatively little attention, beyond a few displaced-vertex analyses of inelastic DM and similar models, e.g. at Belle II~\cite{Duerr:2020muu,Kang:2021oes}. 
As noted in~\cite{Gori:2022vri}, the US DMNI has not yet funded any experiments explicitly optimized for (semi-) visible dark sector searches. Nevertheless, a vibrant program at CERN and other laboratories is beginning to fill this gap. Large datasets from existing collider experiments (e.g. Belle II, LHCb and ATLAS/CMS) can be reexamined with broader, more inclusive analyses designed to spot shorter-lived semi-visible states. 
Many collaborations, including NA62, NA64, Belle II, FASER@LHC, are already testing DS, albeit often under minimal or fully invisible assumptions~\cite{Agrawal:2021dbo,Antel:2023hkf}, see also~\cite{Mongillo:2023hbs,Abdullahi:2023tyk}.
In addition, neutrino detectors (ICARUS, SBND, JUNO) originally built for oscillation physics can act as beam-dump facilities~\cite{Ballett:2016opr,Ballett:2019bgd,Arguelles:2019xgp,Batell:2022xau}, leveraging high-resolution tracking to uncover multi-lepton or partially invisible final states. 
Other proposals --- including $\eta$-factories~\cite{Gan:2015nyc,REDTOP:2022slw,Liu:2024pgf}, electron beam dumps, e.g. LDMX, along with surface and near-detector concepts such as MATHUSLA, and CODEX-b (for a review see \cite{Agrawal:2021dbo,Antel:2023hkf}) --- could expand coverage of exotic new physics if SLP lifetimes and geometries align. Looking ahead, future experiments like SHiP and the DUNE near detectors promise substantially enhanced reach in both lifetime and mass range. The key to exploiting these opportunities lies in integrating robust event generators for multi-particle dark cascades into the analysis pipeline, allowing collaborations to disentangle production and decay channels. Ultimately, combining the complementary capabilities of colliders, fixed targets, and neutrino beam dumps will enable a more comprehensive exploration of MeV--GeV dark sectors, bridging gaps left by purely long-lived or purely invisible searches.

\section{A Prototype for a rich dark sector model: the 3-portal model}

\begin{figure*}[t]
    \centering
    \includegraphics[width=\textwidth]{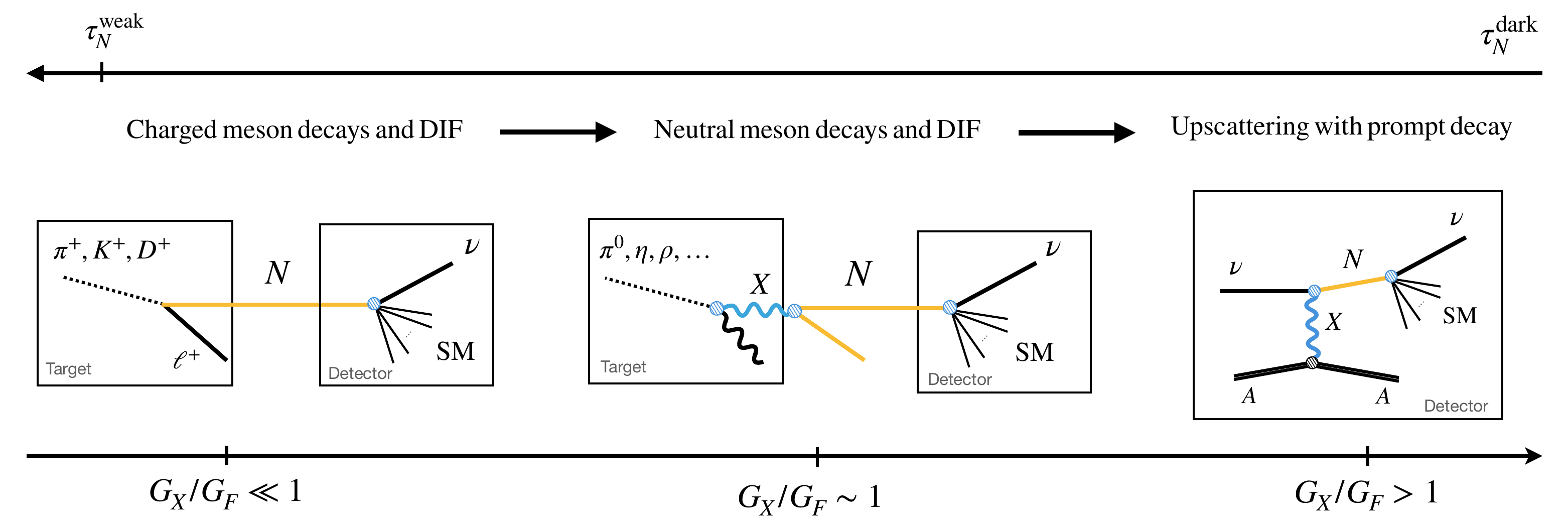}
    \caption{An illustration of the impact of new forces on the signatures of a HNL at a neutrino experiment.
    From left to right, the new force strenght $G_X \sim g_X^2/M_X^2$ increases and the HNL lifetime decreases.
    In the presence of this new mediator, new production channels for HNLs open up, including new meson decays, but eventually, the new particles are short-lived enough that neutrino upscattering becomes the most promising avenue to search for these states.
    \label{fig:darkforcesHNL}
    }
\end{figure*}

While the possibilities for rich dark sector models are vast, to combine brevity and concreteness we focus here on those that have a structure similar to that of the SM, with multiple particles and interactions, as shown in \cref{fig:3portal_examples}.
Within this class of RDS, there is a plethora of possibilities, but a large fraction of them can be represented by prototype models that have key ingredients that capture the typical characteristics and phenomenological signatures of RDS. 
One example is the 3-portal model discussed in~\cite{Ballett:2019pyw} and similar variations~\cite{Harnik:2012ni,Ko:2014bka,Bertuzzo:2018itn,Bertuzzo:2018ftf,Jho:2020jfz}.
We assume that the new physics mass scale is in the MeV-GeV range to avoid big bang nucleosynthesis bounds while remaining well below the EW scale. 
We introduce a new $U(1)$ gauge symmetry with the associated dark photon $X^\mu$, a complex scalar $\Phi$ that spontaneously breaks the $U(1)$ gauge symmetry, and dark fermions, some charged under the symmetry, the $\nu_D$s, and some not, the $N$s.
The Lagrangian reads 
\begin{align} \label{eq:lagrangian}
\mathscr{L}  =  & ~ \mathscr{L}_{\mathrm{SM}} + \left(D_\mu \Phi\right)^\dagger \left(D^\mu \Phi\right) -  V(\Phi,H) \,  - \frac{1}{4}X^{\mu \nu} X_{\mu \nu} 
\\\nonumber
&-\frac{\sin{\chi}}{2} B_{\mu\nu}X^{\mu\nu}+ \overline{N_i}i\slashed{\partial}N_i + \overline{\nu_{D, j}}i\slashed{D}\nu_{D,j} 
\\\nonumber
&- \left[y^\alpha_\nu (\overline{L_\alpha} \cdot \widetilde{H})N^c_i + \overline{N_i}\frac{\mu}{2} N_j^c + y_N \overline{N_i}\nu_{D,j}^c\Phi + \text{h.c.}\right],
\end{align}
where $X^{\mu\nu} \equiv \partial^\mu X^\nu - \partial^\nu X^\nu$, $D_\mu \equiv \left(\partial_\mu-i g_X X_\mu\right)$, $L_\alpha \equiv (\nu_\alpha^T, \ell_\alpha^T)^T$ is the SM leptonic doublet with flavour $\alpha = e, \mu, \tau$ and $\widetilde{H} \equiv i \sigma_2 H^*$ is the conjugate SM Higgs doublet. A kinetic mixing term is allowed with coupling $\sin \chi$; $\epsilon$ is typically used in the literature, $\epsilon = \chi\cos \theta_W $ 
with $\chi \ll 1$ and $\theta_W$ the Weinberg angle. We can introduce any number of $N$ and $\nu_D$ fields, but at least two $N$ fields are required to explain neutrino masses, while the $\nu_D$ fields are chosen to ensure the model is anomaly-free. The gauge coupling of the new symmetry is denoted by $g_X$. The matrices $y_\nu^\alpha$ and $y_N$ are the Yukawa couplings for the $L_\alpha$--$N$ and $\nu_D$--$N$ interactions, respectively.
Since $N$ is neutral under all gauge symmetries, we may include a Majorana mass $\mu$ for $N$. As $N$ carries lepton number, this term is lepton number violating.
The scalar potential includes the SM Higgs doublet and at least one dark scalar so that
\begin{align}
    V(\Phi, H) = &-\frac{\mphi^2}{2} |\Phi|^2 + \frac{\lphi}{4} |\Phi|^4-\frac{\mhiggs^2}{2} H^\dagger H 
    \\\nonumber
    &+ \frac{\lhiggs}{4} (H^\dagger H)^2  + \frac{\lambda_{\phi H}}{4} \, (H^\dagger H)|\Phi|^2 ~.
\end{align}
We assume that the scalar $\Phi$ gets a vacuum expectation value, $v_\Phi$, below the EW scale, due to a negative mass squared term or a negative small coupling to the Higgs, $\lambda_{\phi H}$. 
The vev $v_\Phi$ gives a mass $g_X v_\Phi$ to the dark photon, provides a mass for the dark scalar, and generates a Dirac mass term for $N$ and $\nu_D$, thereby mixing them. In this way, it controls the mass scale of the dark sector, with the exception of the Majorana mass $\mu$. 

This model has a three portal connection to the SM, see \cref{fig:3portal_examples}, as simultaneously a vector, neutrino and scalar portal are present. This allows for multiple windows between the SM and the dark sector and could decouple production and decay mechanisms for the dark particles, changing the bounds obtained with the assumption of minimality.

As lepton number is broken by the Majorana mass of $N$, neutrino masses arise after electroweak and $U(1)^\prime$ symmetry breaking. The $\nu_L$, $N$s, and $\nu_D$ fields will mix and an extended see-saw mechanism will be at play, giving small masses to the standard neutrinos. This will also lead to small mixings $U_{\alpha H}$ between the active SM neutrinos and the heavy neutral lepton states, arising from the diagonalisation of the neutral fermion mass matrix.

Dark matter can be embedded in the model by adding 
a scalar that does not acquire a vev and mixes with neither $H$ nor $\Phi$, or a fermion that does not mix with the neutrinos. 
In the latter case, the stability of the fermionic candidate and the absence of mixing with the neutrinos can be guaranteed by imposing a new symmetry, e.g. a $Z_2$, or a special assignment of the $U(1)^\prime$ charges. As the model has multiple particles that can act as mediators, thermal freeze-out can be easily achieved via annihilations into the SM or into dark particles that subsequently decay to the SM.
Alternatively, a bosonic or fermionic candidate could be introduced with tiny couplings to both the SM and the dark sector via a freeze-in type scenario.

We have taken here a secluded $U(1)^\prime$ under which the SM particles are not charged and $g_X$ is typically large. 
Other anomaly-free options are possible, for example $B-L$ as well as leptonic $L_\alpha-L_\beta$, and provide other interesting examples of RDSM (see, e.g., \cite{Foguel:2024lca} for such models with semi-visible particles). 
It should be noted that in such cases the gauge coupling is constrained to be very small, see e.g.~\cite{Agrawal:2021dbo}.

\begin{figure*}[t]
    \centering
    \includegraphics[width=0.49\textwidth]{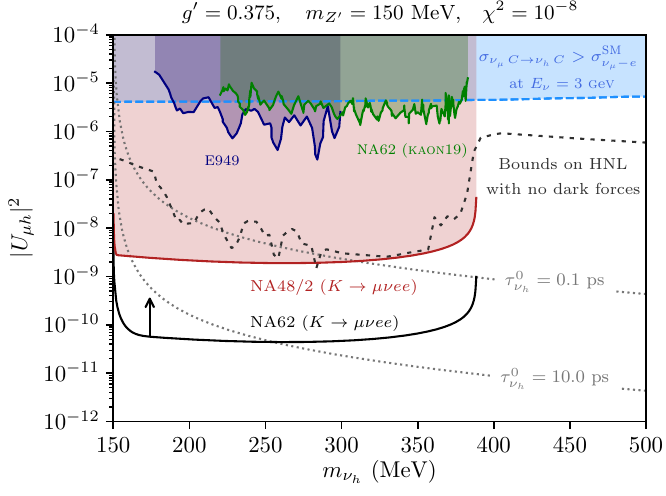}
    \includegraphics[width=0.49\textwidth]{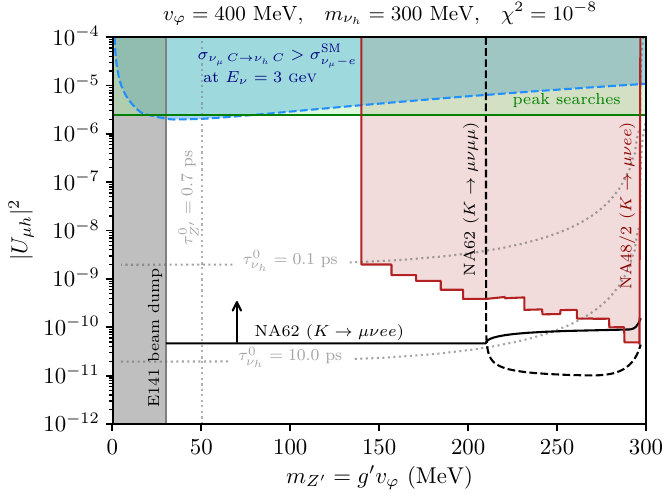}
    \caption{The parameter space of a HNL $\nu_h$ mixing with the muon flavor and coupled to a dark photon that kinematically mixes with the SM photon with $\chi \equiv \varepsilon / c_W = 10^{-4}$. 
    On the left we vary the HNL mass for fixed $m_{Z^\prime} = 150$~MeV and on the right we vary the dark photon mass for a fixed $m_{\nu_h} = 300$~MeV.
    For NA62, we show the zero-background event rate sensitivity.
    Reproduced from \cite{Ballett:2019pyw}.
    \label{fig:HNL_examples}
    }
\end{figure*}

\subsection{Heavy neutral leptons in RDS}
\label{sec:HNL@RDS}

These models typically have multiple neutral fermions at the MeV-GeV scale that mix among themselves and with the light neutrinos. We denote them generically as heavy neutral leptons (HNL). Their phenomenology can be very different from that of minimal models:
\begin{itemize}
    \item \textbf{Minimal models}: the HNLs can decay only into SM particles, via multiple channels e.g. into 3 neutrinos, into electron-positron pairs and neutrinos, into pions and neutrinos or charged leptons, and the decay rate is controlled by $G_F$ times a tiny mixing parameter, so that their lifetime is very long. 
    The branching ratios are set by the SM gauge couplings and kinematics.
    \item \textbf{Rich dark sectors}: the heavier HNLs will decay into lighter ones or into lighter ones and SM particles or into lighter HNLs/neutrinos and the dark scalar, depending on the hierarchy of masses. Such decays will proceed fast since the gauge and Yukawa couplings in the dark sector are not small. The lightest HNL can decay only into SM particles or into neutrinos and dark scalar, depending on the channels that are kinematically open. Also in this case, the decays can be much faster than in the minimal case as the mediators, e.g. the dark photon and the dark scalar, can be light.
\end{itemize}

The fact that HNLs decay much faster than in the minimal case has striking consequences on their searches.
We illustrate this point in \cref{fig:darkforcesHNL}, showing how the presence of a new force changes the lifetime and the lamppost signature of HNLs.
Examples of how experimental search strategies have to be modified are provided below:
\begin{itemize}
    \item Decay in flight: decay searches in beam-dump experiments need to be reconsidered. If the lifetime is much shorter than the distance between the target and the detector, the bounds will not apply as the HNLs will have decayed before reaching the detector, leaving no possible signal. The converse is also possible if the lifetime is much shorter than in the minimal case but not enough to have them decay too fast. In this case, the bounds on the mixing angles will shift to lower values, testing the see-saw preferred-region and will not apply for large values.
    \item Meson decays: pion and kaon decays provide some of the strongest limits on the minimal HNL models, being some of the only searches to touch the region of interest of a minimal Type-I seesaw at $\mathcal{O}(100)$~MeV HNL masses.
    These searches target an invisible resonance in $\pi^+(K^+) \to \ell^+ N_{\rm inv}$ by searching for a peak in the invariant mass $(p_{\pi(K)} - p_\ell)^2 = M_N^2$.
    If the HNL decays (semi-visibly) within the detector, the vetoes on additional activity can weaken these limits.
    In that case, looking for new resonances in multi-lepton pion and kaon decays such as $K^+ \to \mu^+ (N_{\rm dark} \to \nu_\ell (Z^\prime \to e^+ e^-))$ would not only recover the lost sensitivity, but potentially boost it to even lower mixing parameters. 
    For a light mediator produced in the 2-body decays of the HNL, the experimental signature constitutes a striking peak in the 2-dimensional mass plane of $(p_{K} - p_\ell)^2 = M_N^2$ versus $(p_{e^+} + p_{e^-})^2 = M_{Z^\prime}^2$.
    
    In \cref{fig:HNL_examples}, we show the recast of the $K^+\to \mu^+ \nu_\mu e^+e^-$ measurement at NA48/2~\cite{Peruzzo:2017qis} (see also~\cite{Poblaguev:2002ug}) onto the parameter space of a HNL coupled to a dark photon.
    Unfortunately, in both cases, the $e^+e^-$ invariant mass is restricted to be greater than $140$~MeV, so no bound can be derived for $m_{Z^\prime} < 140$~MeV.
    More recently, NA62 has performed the first measurement of $K^+ \to \mu^+ \mu^+ \mu^- \nu_\mu$~\cite{Boretto:2020lun}, which can also be translated into a bound in the same parameter space.
    
    \item Lepton decays: muon and tau decays can also provide relevant limits on the mixing of HNLs to SM particles. Lepton flavor universality in tau decays sets stringent limits on the mixing parameters in a largely-model-independent way (see, e.g., \cite{deGouvea:2015euy,Fernandez-Martinez:2016lgt}).
    For short-lived HNLs, multilepton final states become a promising search strategy.
    The Mu3e experiment could reach mixing parameters as small as $|U_{\mu 4}|^2 \gtrsim 10^{-13}$ by looking for displaced $\mu^+ \to e^+ \nu_e (N \to \overline\nu_\mu e^+e^-)$ decays~\cite{Knapen:2024fvh}.
    One can also envision constraints from analogous tau decays based on searches for tau decays to three charged tracks. 
    \item Upscattering: in minimal models, neutrino-scattering production of HNLs is possible, but extremely rare because of the small number of neutrino interactions in neutrino detectors.
    With a new force that is ``stronger-than-Weak'' ($G_X > G_F)$, the cross section is enhanced and the number of events can be large for tiny mixing angles.
    This is usually accompanied by a subsequent short-lived decay of the HNL to SM particles via the new force, further boosting detection prospects. 
    This signature is the most promising for MeV-GeV dark mediators coupled to neutrinos and HNLs.
    Accelerator experiments are well-suited to search for this because the HNL mass imposes a threshold in the cross section. The heavier the HNL, the higher energy neutrino beam is required.
    
\end{itemize}

New production and decay modes of HNLs in RDS have been considered in the context of axion-like-partilces~\cite{Chang:2021myh,Abdullahi:2023gdj,Li:2023dbs,Wang:2024prt}, gauge symmetries~\cite{Jho:2020jfz,Batell:2016zod,Capozzi:2024pmh}, transition magnetic moments~\cite{Magill:2018jla,Fischer:2019fbw,Brdar:2020quo,Bertuzzo:2024eds}, and other effective operators~\cite{Duarte:2015iba,Liao:2016qyd,Delgado:2022fea,Fernandez-Martinez:2023phj}.
In many of these cases, the HNL parameter space becomes cosmologically viable below the kaon mass, thanks to the much shorter lifetimes~\cite{Arguelles:2021dqn}.

If the dark sector fermions have negligible mixing with neutrinos (or identically zero), they can constitute dark matter.
With multiple generations, the dark fermions decay in cascades down to the lightest stable particle, i.e., the dark matter candidate.
Pairing the dark fermions into pseudo-Dirac pairs, leads to the well-known inelastic DM scenario~\cite{Tucker-Smith:2001myb}: dark matter has no self-interactions, but can be produced via freeze-out through co-annihilations, $\chi_1 \chi_2 \to {\rm SM}\,{\rm SM}$.
In this case, the phenomenology can be somewhat similar to that of the HNLs discussed above.
The main differences include the new production and decay modes that are controlled by mixing with neutrinos and the relaxation of the requirement to reproduce the DM relic abundance.

\subsection{Dark photons in RDS}
\label{sec:DP@RDS} 

\begin{figure*}[t]
    \centering
    \includegraphics[width=0.49\textwidth]{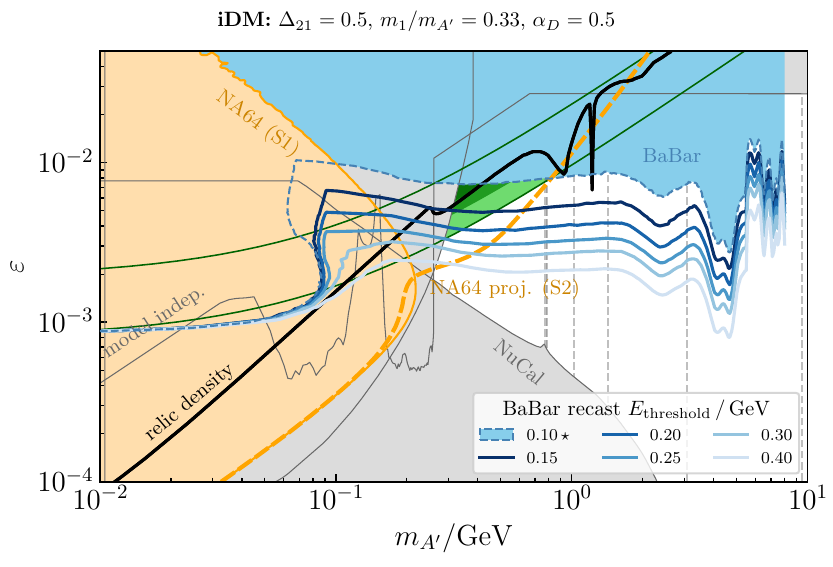}
    \includegraphics[width=0.49\textwidth]{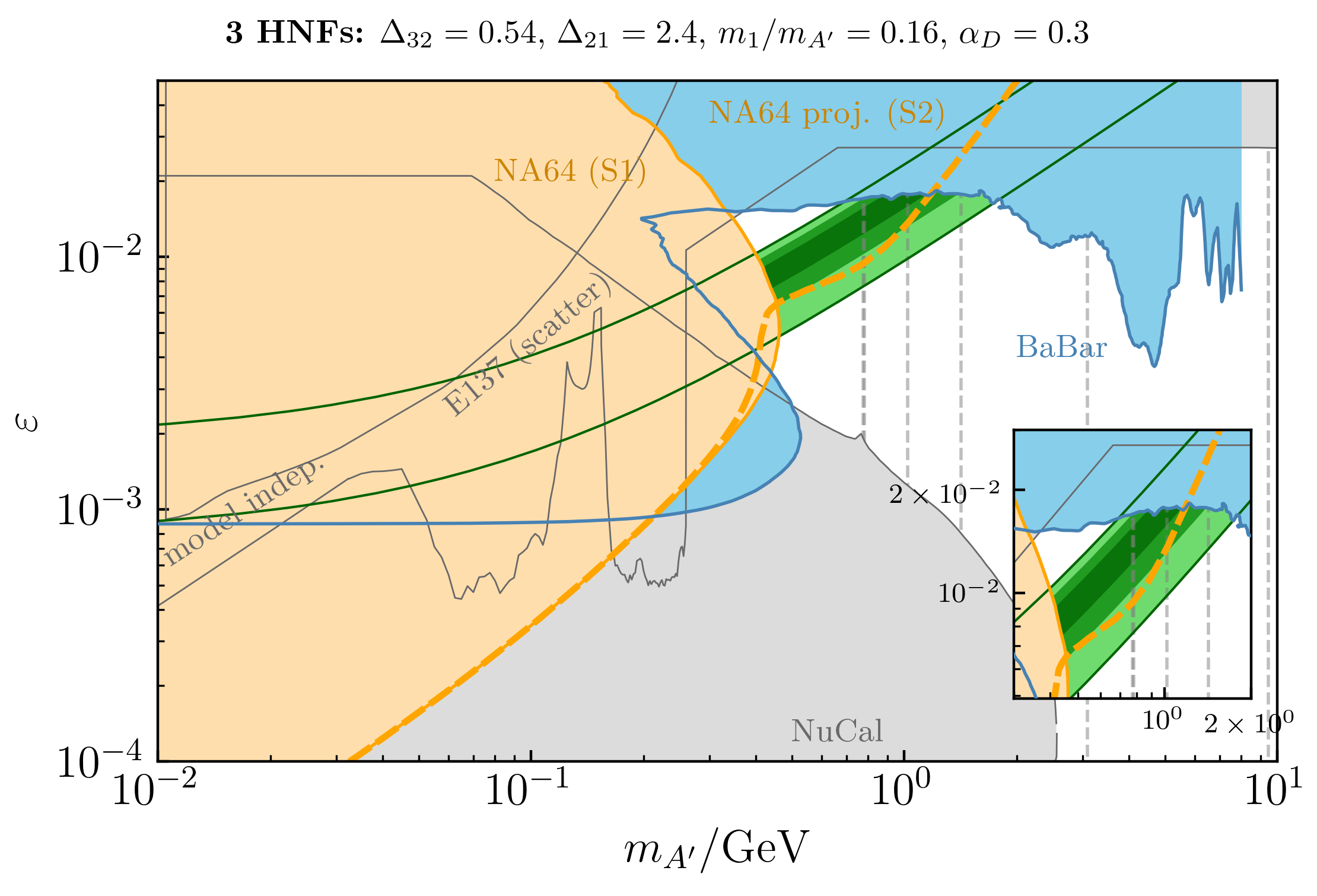}
    \caption{Slices of the parmeter space of semi-visible dark photons; on the left, we show the recast of existing limits on inelastic dark matter. On the right, we show the same, but for a model with three heavy neutral fermions (HNF), which may or may not mix with neutrinos.
    The various shades of blue in the BaBar recasted limits represent different assumptions on the veto thresholds of the original analysis.
    Reproduced from \cite{Abdullahi:2023tyk}.
    \label{fig:darkphoton_examples}
    }
\end{figure*}

In these models dark photons can have multiple interactions: the ones with dark particles charged under the new symmetry will be controlled by the $g_X$ coupling that can be expected to be quite large as for the SM gauge couplings if finetuning is not invoked; the ones with the SM are instead suppressed by small couplings and are subdominant. This results in typical decays of the dark photons into dark particles if kinematically allowed. The hierarchy of dark particle masses depends on the choice of the scalar, gauge and Yukawa couplings and so on the specific model. In the most naive case, one could expect a similar structure as to the SM, with dark photon and scalar boson quite heavy and dark fermions a bit lighter. 
This has important phenomenological consequences that make dark photon searches potentially very different to the case of minimal models.

While a higgsed dark photon model has received some attention in the literature (see, e.g., recent studies in ~\cite{Foguel:2022unm}), dark photons and scalars coupled to  dark fermions has been largely overlooked, with the exception of the inelastic DM regime discussed above.
In what follows, we outline some of the interesting signatures of dark photons in RDSM.

\begin{itemize}
    \item {\bf Minimal models}: the dark photon decays into SM via kinetic mixing. If it is sufficiently light, it will decay only into dilepton pairs, in particular electron-positrons. This decay is purely visible with no missing energy and experiments typically impose cuts on the missing energy to reduce the SM backgrounds. The decay length can be significant as the kinetic mixing coupling is very small and the typical signature is that of displaced vertices, as appropriate for LLP searches.
    
    Some minimal extensions of these models include DM candidates and dark photons could dominantly decay into a pair of DM particles, if the latter are sufficiently light. In this case, the dark photon decay is purely invisible as the DM particles exit the detector. In this case, cuts on visible energy are put to avoid large SM backgrounds. The decay could be fast but this does not impact the experimental search strategy.

    \item {\bf Rich dark sectors}: dark photons will typically decay fast into lighter dark fermions, if present. The latter will further decay as discussed in the Section above, leading to semivisible or invisible decays. The most interesting case is that of dark fermions that decay into SM and missing energy as the dark photon signature would be of a fast semivisible decay. The cuts discussed above for visible and invisible searches would hide the dark photon signature into the background sample.
\end{itemize}

Extensive searches for dark photons have been done with strong bounds on kinetic mixing in both the short-lived and long-lived $Z^\prime$ regime.
In non-minimal models, many of them become irrelevant and one should consider instead searches for multi-leptons such as those discussed below.
\begin{itemize}
    \item Searches at $e^+e^-$ colliders: invisible dark photons are effectively constrained by searches for $e^+e^- \to \gamma Z'$, that is, single photons recoiling against missing energy.
    With a very small subset of their total luminosity, BaBar sets limits at the level of $\epsilon < 10^{-3}$~\cite{BaBar:2017tiz} with this approach. 
    For a semi-visible dark photon, the constraints can be weakened and new multi-lepton signatures become more promising~\cite{Mohlabeng:2019vrz,Duerr:2019dmv,Abdullahi:2023tyk}.
    We show how the BaBar constraints can change depending on the dark fermion content of the dark sector in \cref{fig:darkphoton_examples}.
    To cover the new gaps in parameter space, direct production of the dark fermions with or without radiative return can be pursued at future experiments like Belle-II~\cite{Duerr:2020muu,Kang:2021oes,Acevedo:2021wiq,Ko:2025drr}.   
    
    When the dark photon is produced alongside the dark higgs in ``dark higgsstrahlung", $e^+e^- \to Z' h'$, the multiplicity of leptons in the final state can quickly multiply when the dark higgs decays as $h' \to Z'Z' \to 2(\ell^+\ell^-)$.
    By searching for three visible resonances,  BaBar~\cite{BaBar:2012bkw} and Belle~\cite{Jaegle:2015fme} have placed strong limits on the parameter space of higgsed dark photons.
    More recently, Belle-II~\cite{Belle-II:2022jyy} has expanded this reach to scenarios where the dark higgs is long-lived or invisible.
    For a recent overview of dark higgs searches at colliders, see~\cite{Ferber:2023iso}.

    \item Searches at the LHC: at the energy frontier, it is possible to source dark particles from the decays of the SM higgs.
    This provides a unique method to access the RDSM (see, e.g., ~\cite{Falkowski:2010cm,Falkowski:2010gv}) and provides a direct probe of the parameters in the scalar potential.
    Most notably, $H \to h' \to Z' Z'$ leads to boosted multi-lepton signatures that have received increased attention in ATLAS~\cite{ATLAS:2018rjc,ATLAS:2021ldb,ATLAS:2022izj,ATLAS:2024zxk} and CMS searches~\cite{CMS:2012qms,CMS:2015nay,CMS:2021pcy,CMS:2021sch,CMS:2024qxz}.
    Higgs decays can also produce meta-stable semi-visible fermions in the RDSM, which can be searched for at the lifetime frontier of the LHC~\cite{Li:2021rzt}. 
    
    \item Meson decays: light dark photons are effectively constrained by searches for visible resonances in $\pi^0 \to \gamma Z^\prime$ decays~\cite{NA482:2015wmo}.
    Through a cascade of decays, dark photons can also be produced in meson meson decays such as $K\to h'\to Z'Z'$ and $K\to \pi ( h'\to Z'Z')$ or $B \to h' \to Z'Z'$ and $B \to K (h' \to Z'Z')$ via the mixing between the higgs and the dark higgs~\cite{Batell:2009jf,Hostert:2020xku,Cheung:2024oxh}.
    The $K_L \to 2(e^+e^-)$ decays have been measured by KTEV~\cite{KTeV:2001nui} and NA48~\cite{NA48:2000szm} (searches for $K_L \to XX \to 4 \gamma$ have also been performed by KOTO recently~\cite{KOTO:2022lxx}), but no such measurement exists for $K_S$.
    A constraint on $K_S \to 2(\mu^+\mu^-)$ was recently imposed by LHCb~\cite{LHCb:2022tpr} and searches for $B$ decays to four leptons have been performed by Belle~\cite{Belle:2020the}.
    Searches for five charged tracks in kaon decays have been performed for the first time with $K^+ \to \pi^+ 2(e^+e^-)$ by NA62~\cite{NA62:2023rvm} and with $K_L\to\pi^0 2(e^+e^-)$ by KOTO~\cite{Li:2024ieq}.
    All aforementioned channels set limits on $\theta^2_{h h'}$, as opposed to limits on $(\epsilon g_D)^2$ from dark higgsstrahlung at $e^+e^-$ colliders discussed above.
    
    \item Searches at beam dumps: the decay in flight of the dark photon is typically fast, but can lead to the production of metastable, semi-visible dark states (i.e., the products of their decays contain missing energy and visible particles). 
    The couplings controlling production and the lifetime of the metastable dark partners are in general different. 
    Because decays are generally faster in these models, beam dumps with a small baseline and large acceptance like SHiP could set strong bounds in the future.
    Much of the literature in this case has focused on inelastic DM models.
    For a non-exhaustive list of beam dump constraints and sensitivity studies, see, e.g.,~\cite{Izaguirre:2015zva,Izaguirre:2017bqb,Berlin:2018jbm,Tsai:2019buq,Jodlowski:2019ycu,Batell:2021ooj,Jodlowski:2021xye,Garcia:2024uwf}.
        
    \item Fixed targets: the presence of displaced vertices and the semi-visible decays of the dark photon can interfere with the search for missing energy at fixed target experiments like NA64~\cite{NA64:2016oww} and LDMX~\cite{LDMX:2018cma}.
    In this case, new strategies dedicated to finding the displaced activity can improve the sensitivity to RDS.
    For example, NA64 has constrained semi-visible dark photons with both a dedicated displaced vertex search~\cite{NA64:2021acr} and with a rescast of the invisible dark photons searches~\cite{Mongillo:2023hbs}.
\end{itemize}

\begin{figure*}[t]
    \includegraphics[width=0.49\textwidth]{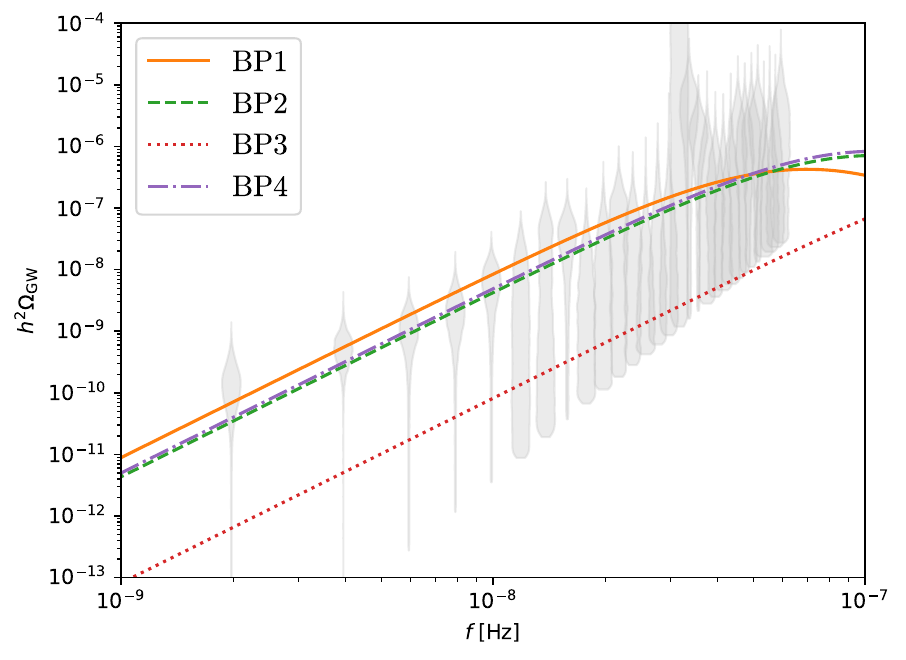}
    \includegraphics[width=0.49\textwidth]{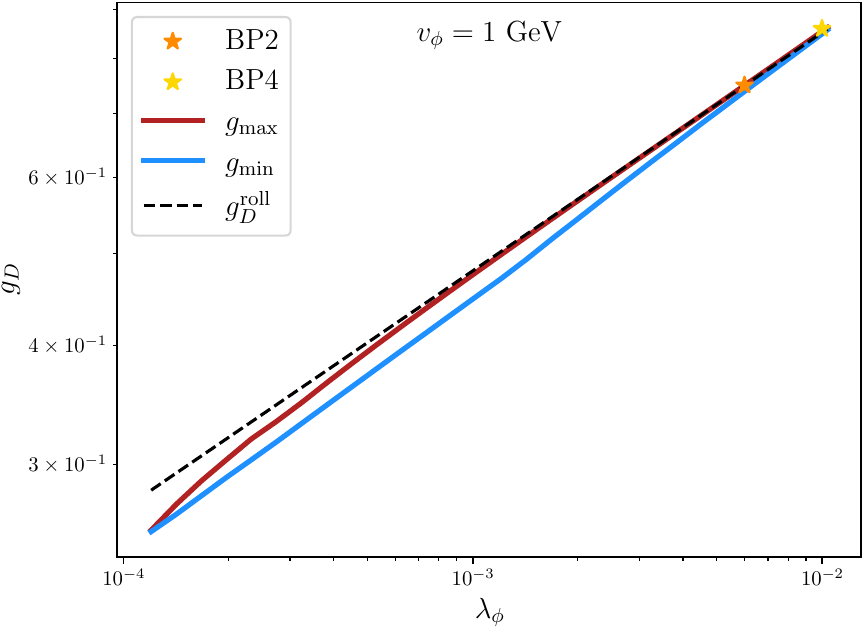}
    \caption{
    Left: the SGWB generated by a FOPT in the model in Eq.~(\ref{eq:pta_model}) for selected benchmark points.
    Right: the maximum (minimum) value $g_\mathrm{max}$ ($g_\mathrm{min}$) of the gauge coupling $g_D$ for which a FOPT completes in the runaway regime, as a function of the scalar quartic coupling $\lambda_{\phi}$. The dashed line corresponds to the relation in Eq.~(\ref{eq:g_roll}). Further details can be found in~\cite{Costa:2025csj}.
  \label{fig:pta_model}}
\end{figure*}


\subsection{Dark Scalars and phase transitions in the Early Universe}

Unlike in the SM, the interaction between the scalar and gauge sectors in RDS scenarios can introduce richer dynamics in the Early Universe. In fact, when the gauge symmetry in the DS is spontaneously broken, it can lead to a FOPT. A sufficiently supercooled FOPT generates a strong GW signal that PTA observatories can probe. A DS FOPT is one of the most compelling new physics explanations for the NANOGrav Stochastic GW Background (SGWB) observation~\cite{Fujikura:2023lkn,Addazi:2023jvg,Conaci:2024tlc,Banik:2024zwj,Goncalves:2025uwh,Costa:2025csj,Balan:2025uke}. A signal in the nanohertz frequencies points toward a (sub-)GeV DS vev~\cite{Winkler:2024olr,Ghosh:2023aum,Bringmann:2023opz,Costa:2025csj,Balan:2025uke}.

An example~\cite{Costa:2025csj} is given by a low-scale DS composed of a complex scalar $\phi$ charged under a dark $U(1)_\mathrm{D}$ gauge symmetry, and the associated dark gauge boson $Z^{\prime}_{\mu}$ 
\begin{align}\label{eq:pta_model}
    \mathcal{L} = &\left(D_{\mu}\phi\right)^{*}\left(D^{\mu}\phi\right) -\frac{1}{4}Z^{\prime}_{\mu\nu}Z^{\prime \mu\nu} -V(\phi^*\phi),
\end{align}
with $V=-\mu_{\phi}^2\phi^*\phi+\lambda_{\phi}\left(\phi^*\phi\right)^2$.
Assuming $\mu_{\phi}^2 > 0$, the dark scalar breaks the symmetry by acquiring a vev $v_\phi=\mu_{\phi} / \sqrt{\lambda_\phi}$, resulting in the mass spectrum $m_{Z'}^2 = g_D^2 v_\phi^2$ and $m_{\phi}^2 = 2\lambda_{\phi} v_\phi^2$.

While the frequency of the GW background observed by NANOGrav sets the scale of new physics $v_\phi$ to be around the (sub-)GeV scale, the fact that its duration is long, i.e. it is supercooled, calls for a very particular shape in the potential, resulting in a nearly conformal scalar sector. This is realised if the gauge coupling lies close to the line
\begin{equation}
g_{D} = \left\lbrace \frac{16\pi^2\lambda_{\phi}}{3}\left[1-\frac{\lambda_{\phi}}{8\pi^2}\left(5+2\log{2}\right)\right]\right\rbrace^{1/4}\,.
\label{eq:g_roll}
\end{equation}
The relation above has the strong phenomenological implication of a specific relation of masses, {$m_\phi^2 / m_{Z'}^2 \simeq \sqrt{3 \lambda_\phi} / 2 \pi$}, that can be targeted and tested in experimental searches.
We show in Fig.~\ref{fig:pta_model}-left the SGWB generated by selected benchmark points, see Ref.~\cite{Costa:2025csj},  in the model in Eq.~(\ref{eq:pta_model}) compared to the periodogram inferred by the NANOGrav collaboration, while on  Fig.~\ref{fig:pta_model}-right we highlight the region of parameter space where phenomenologically viable FOPT are possible, compared to the line in Eq.~(\ref{eq:g_roll}). Further details are available in~\cite{Costa:2025csj}.

Once the PT is completed, the DS energy density needs to be transferred to the visible sector before the onset of big bang nucleosynthesis. In fact, it has been shown that stable DSs are in tension with NANOGrav data \cite{Bringmann:2023opz}. In RDS, the energy transfer is achievable via the Higgs or the neutrino portals, which allow the scalar to decay sufficiently fast into the SM. This requirement sets a lower bound on the strength of the coupling constants and can suggest the correct DS parameter space region to probe in terrestrial experiments.

\section{Strategies for bounds on dark particles, beyond LLPs}

Current bounds (e.g. see \cite{Agrawal:2021dbo,Antel:2023hkf}) on dark particles rely on the assumption of minimality. As discussed above, these bounds are {\bf model-dependent} as they may not apply if these particles decay via semivisible channels and/or fast. Abandoning minimality opens a vast landscape of models, and it is essential that experimental bounds can be applied to multiple models or can be easily recasted.  We would like to present some suggestions to facilitate such effort.
\begin{itemize}
    \item It is useful to focus on the physical properties of the dark particles, namely masses, decay length and scattering cross sections. In this way, bounds constrain measurable quantities, that can be used to test many models and in other experiments.\footnote{Similar points have been made for searches for long-lived particles~\cite{Batell:2023mdn}.}
    \item Searches should be done as inclusively as possible to capture a broad range of possible signatures that can arise in multiple models. In addition to this type of searches, the experimental collaborations could also carry out dedicated analysis for specific benchmark models, to obtain more stringent bounds, although model-dependent.
    \item Dark particle production and decay should be decoupled as they may be induced by different interactions. Constraints on portal couplings, should account for different assumptions on the particle lifetime and/or production cross section. 
    \item The assumptions on the models should be clearly stated so that the bounds could be easily recasted. To this aim, the expression for the dominant decay rates, production cross sections with their energy and angular dependences, should be given for each bound.
    \item If the events depend on multiple interactions, several benchmark points should be chosen, so that they capture the key behaviours of the events: for instance, heavy and light mediators, two-body versus three-body decays, compressed fermionic spectra versus large mass hierarchies. 
    \item As rich dark sectors have signatures in multiple type of experiments, various searches need to be combined, together with information from astroparticle physics and cosmology, to constrain the models and to identify the most promising benchmark points. This allows to achieve a physics reach vastly superior to that of any individual experiments. 
    \item Collaboration with phenomenologists and astroparticle theorists should be strongly encouraged to consider particularly interesting and promising models, to identify the allowed parameter space and to take into account implications and constraints from astroparticle physics and cosmology.
\end{itemize}

We stress that astrophysical probes of RDS provide an important and complementary effort to laboratory ones. 
In fact, several puzzles in ultra-high-energy neutrino data are currently observed.
These include the ultra-high-energy neutrino-like events at ANITA~\cite{ANITA:2018sgj,ANITA:2020gmv} and KM3NeT~\cite{KM3NeT:2025npi}, which defy astrophysical explanations based on standard neutrino emission by astrophysical sources.
See, for instance, \cite{Safa:2019ege} and \cite{Li:2025tqf} for a discussion of these observations in the context of IceCube data.
Another interesting example is the capture of dark matter in celestial bodies, neutron stars, or white dwarfs. 
In the last years, there has been an effort to describe these processes with sufficient precision~\cite{Kouvaris:2010vv, Baryakhtar:2017dbj, Garani:2018kkd, Bell:2020jou, Bell:2020lmm, Anzuini:2021lnv, Bramante:2023djs, Bell:2023ysh, HoefkenZink:2024hor}, so that the data taken by telescopes such as the FAST radio telescope, the James Webb Space Telescope (JWST), Thirty Meter Telescope (TMT), and European Extremely Large Telescope (E-ELT) can be used to constrain dark matter interactions with particles from the SM. White dwarf cooling could also point to dark hidden sectors, although their apparent inexistent deviation from SM predictions~\cite{Hansen:2015lqa} serve as a tool to constrain DS parameter space~\cite{Zink:2023szx, Foldenauer:2024cdp}.

\section{Summary and conclusions}

In the search for a New Standard Model theory, that explains the evidence of physics beyond the SM and the fundamental properties of the SM, recently there has been a broadening of the physics scales under scrutiny, thanks also to theoretical developments and new experimental opportunities. In particular, dark sectors, i.e. extensions of the SM below the electroweak scale, have seen a blooming of interest and a strong experimental programme with many new searches. The dark sector can interact with the SM via renormalizable terms with tiny couplings, namely the vector, scalar and neutrino portal, and/or via effective field theory operators arising from some higher scale theory. Possible hints may have been found but require further experimental scrutiny to establish their validity.

Most of the studies, and nearly all experimental searches, have been done under the assumption of minimality, that is adding to the SM the smallest number of ingredients, typically just one particle and its portal to the SM, plus dark matter in some cases. For comparison, the SM is a highly non minimal theory with a gauge sector, a scalar doublet, needed to break the electroweak symmetry and setting the mass scale of the theory, and a plethora of fermions of different types and in three generations. A dark sector could have a similar complex structure, hence its denomination as ``rich": it can have new symmetries, gauged and/or local, new dark scalars to break (some) of them and give raise to the new energy scale, and multiple fermions. Some of these particles could also make the DM of the Universe. This type of models have many theoretical implications, e.g. for the explanation of neutrino masses, the dark matter and maybe the baryon asymmetry of the Universe. 
They can have a phenomenology very different from minimal models. Due to the multiple dark particles and non-negligible interactions among them, typically the dark particles decay fast, being shortlived, rather than long lived as in minimal models, and have different decay channels.
Searches of minimal heavy neutral leptons, dark photons, dark scalars have been done reaching strong sensitivities but could have missed these signatures due to the experimental configurations and analysis strategy. Heavy neutral leptons can decay very fast into other dark particle and/or SM, requiring a reanalysis of decay-in-flight searches and novel experimental strategies such as upscattering in neutrino experiments; dark photons/scalars would decay fast semivisibly, instead of purely visibly/invisibly as in minimal models; dark matter can have multiple components and significant interactions with dark/SM particles with consequences for their production, detection and impact on the evolution of the Universe. We have reviewed some notable examples of rich dark sector models at the MeV--GeV scale with their phenomenology. 

We advocate a dedicated programme of searches for rich dark sectors that overcomes the assumptions on minimality and
on the long lifetime of particles and encompasses a broader range of possibilities. A combined effort between theorists and experimentalists is needed to explore these possibilities and fully exploit the wealth of present and future experimental opportunities.

\section*{Acknowledgments}
This work is based on an extended revision of the Input submitted by some of the authors to the European Strategy for Particle Physics - 2026 update (ID\#: 250).
M.~L. thanks Fermilab for hosting him during the development of this document. M.~L. is funded by the European Union under the Horizon Europe's Marie Sklodowska-Curie project 101068791 — NuBridge. F. C. acknowledges partial support from the FORTE
project CZ.02.01.01/00/22 008/0004632 co-funded by
the EU and the Ministry of Education, Youth and
Sports of the Czech Republic. J.~H.~Z. is supported by the National Science Centre, Poland (research grant No. 2021/42/E/ST2/00031). A.~G.~D.~M. thanks Lawrence Berkeley National Laboratory for kind hospitality during the developement of this document. The research reported has received partial support from the European Union’s Horizon 2020 research
and innovation programme under the Marie Sklodowska-Curie grant agreement No~860881-HIDDeN and and No.~101086085-ASYMMETRY, by COST (European Cooperation in Science and Technology) via the COST Action COSMIC WISPers CA21106 and by the Italian INFN program on Theoretical Astroparticle Physics (TAsP).

\bibliographystyle{elsarticle-num}
\bibliography{main}

\end{document}